\documentclass[12pt,a4paper]{report}

\usepackage{thesis}
\usepackage{rotating}
\usepackage{lmodern}
\usepackage[numbers,sort&compress]{natbib}
\usepackage[hidelinks]{hyperref}
\usepackage[left=2.5cm,top=2.5cm,right=2.5cm,bottom=2.5cm]{geometry}
\usepackage{nomencl}
\usepackage{amsmath,amssymb,amsfonts}
\usepackage{algorithmic}
\usepackage{graphicx}
\usepackage{textcomp}
\usepackage{xcolor}
\usepackage{mathtools}
\usepackage{float}
\usepackage{amsthm}

\usepackage{listings, xcolor}

\makenomenclature

\usepackage{tocloft}

\newcommand{\thesistitle}{Lazy Contracts: Alleviating High Gas Costs by Secure and Trustless Off-chain Execution of Smart Contracts}
\newcommand{\thesisauthor}{Soroush Farokhnia}
\newcommand{\programname}{Computer Science and Engineering}
\newcommand{\departmentname}{Department of Computer Science and Engineering}
\newcommand{\thesisdate}{June 2023}
\newcommand{\signdate}{}
\newcommand{\thesisdegree}{Master of Philosophy (MPhil)}
\newcommand{\dedicate}{\textit{To Sepehr.}}
\newcommand{\supervisorinfo}{Prof. Amir Kafshdar Goharshady\\ Thesis Supervisor \\ Departments of Computer Science and Mathematics}

\newcommand{\depheadinfo}{Prof. Xiaofang Zhou \\ Department Head\\ Department of Computer Science and Engineering}
\usepackage{enumerate}

\usepackage{listings, xcolor}

\definecolor{verylightgray}{rgb}{.97,.97,.97}

\lstdefinelanguage{Solidity}{
	keywords=[1]{anonymous, assembly, assert, balance, break, call, callcode, case, catch, class, constant, continue, constructor, contract, debugger, default, delegatecall, delete, do, else, emit, event, experimental, export, external, false, finally, for, function, gas, if, implements, import, in, indexed, instanceof, interface, internal, is, length, library, log0, log1, log2, log3, log4, memory, modifier, new, payable, pragma, private, protected, public, pure, push, require, return, returns, revert, selfdestruct, send, solidity, storage, struct, suicide, super, switch, then, this, throw, transfer, true, try, typeof, using, value, view, while, with, addmod, ecrecover, keccak256, mulmod, ripemd160, sha256, sha3}, 
	keywordstyle=[1]\color{blue}\bfseries,
	keywords=[2]{address, bool, byte, bytes, bytes1, bytes2, bytes3, bytes4, bytes5, bytes6, bytes7, bytes8, bytes9, bytes10, bytes11, bytes12, bytes13, bytes14, bytes15, bytes16, bytes17, bytes18, bytes19, bytes20, bytes21, bytes22, bytes23, bytes24, bytes25, bytes26, bytes27, bytes28, bytes29, bytes30, bytes31, bytes32, enum, int, int8, int16, int24, int32, int40, int48, int56, int64, int72, int80, int88, int96, int104, int112, int120, int128, int136, int144, int152, int160, int168, int176, int184, int192, int200, int208, int216, int224, int232, int240, int248, int256, mapping, string, uint, uint8, uint16, uint24, uint32, uint40, uint48, uint56, uint64, uint72, uint80, uint88, uint96, uint104, uint112, uint120, uint128, uint136, uint144, uint152, uint160, uint168, uint176, uint184, uint192, uint200, uint208, uint216, uint224, uint232, uint240, uint248, uint256, var, void, ether, finney, szabo, wei, days, hours, minutes, seconds, weeks, years},	
	keywordstyle=[2]\color{teal}\bfseries,
	keywords=[3]{block, blockhash, coinbase, difficulty, gaslimit, number, timestamp, msg, data, gas, sender, sig, value, now, tx, gasprice, origin},	
	keywordstyle=[3]\color{violet}\bfseries,
	identifierstyle=\color{black},
	sensitive=false,
	comment=[l]{//},
	morecomment=[s]{/*}{*/},
	commentstyle=\color{gray}\ttfamily,
	stringstyle=\color{red}\ttfamily,
	morestring=[b]',
	morestring=[b]"
}

\newcommand{\contract}{\mathcal{C}}
\newcommand{\wrapped}{\mathcal{L}}
\newcommand{\parties}{\mathcal{P}}

\newcommand{\alice}{\textsc{Alice}}
\newcommand{\bob}{\textsc{Bob}}
\newcommand{\ingrid}{\textsc{Ingrid}}
\newcommand{\paul}{\textsc{Paul}}
\renewcommand{\gamma}{gb}

\newcommand{\keyword}[1]{\texttt{\textbf{\textcolor{blue}{#1}}}}
\newcommand{\gv}[1]{\texttt{\textbf{\textcolor{violet}{#1}}}}

\newcommand{\todo}{\textcolor{red}{\textbf{TODO }}}
\renewcommand{\todo}[1]{#1}
\renewcommand{\paragraph}[1]{\smallskip\noindent\textbf{{#1.}}}

\theoremstyle{definition}

\begin{document}

\pagenumbering{roman}
\pagestyle{plain}
\setcounter{page}{1}
\addcontentsline{toc}{chapter}{Title Page}
\thispagestyle{empty}
\null\vskip0.5in
\begin{center}
  \begin{LARGE}
    \thesistitle
  \end{LARGE}
  \vfill
  \vspace{20mm}

  by

  \vspace{4mm}

  \thesisauthor \\
  \vfill
  \vspace{20mm}

  A Thesis Submitted to\\
  The Hong Kong University of Science and Technology \\
  in Partial Fulfillment of the Requirements for\\
  the Master of Philosophy (MPhil) Degree \\
  in \programname \\
  \vfill \vfill
  \thesisdate, Hong Kong
  \vfill
\end{center}

\vfill

\newpage
\addcontentsline{toc}{chapter}{Authorization Page}
\null\skip0.2in
\begin{center}
{\bf \Large \underline{Authorization}}
\end{center}
\vspace{12mm}

I hereby declare that I am the sole author of this thesis.

\vspace{10mm}

I authorize the Hong Kong University of Science and Technology to lend this
thesis to other institutions or individuals for the purpose of scholarly research.

\vspace{10mm}

I further authorize the Hong Kong University of Science and Technology to
reproduce the thesis by photocopying or by other means, in total or in part, at the
request of other institutions or individuals for the purpose of scholarly research.

\vspace{30mm}

\begin{center}
\underline{~~~~~~~~~~~~~~~~~~~~~~~~~~~~~~~~~~~~~~~~~~~~~~~~~~~~~~~~~~~~~~~~~~~~~~}\\
~~~~\thesisauthor \\
~~~~\signdate

\end{center}

\newpage
\addcontentsline{toc}{chapter}{Signature Page}
\begin{center}
{\Large \thesistitle}\\
\vspace{5mm}
by\\
\vspace{3mm}
\thesisauthor\\
\vspace{5mm}
This is to certify that I have examined the above MPhil thesis\\
and have found that it is complete and satisfactory in all respects,\\
and that any and all revisions required by\\
the thesis examination committee have been made.
\end{center}

\vspace{15mm}

\begin{center}
\underline{~~~~~~~~~~~~~~~~~~~~~~~~~~~~~~~~~~~~~~~~~~~~~~~~~~~~~~~~~~~~~~~~~~~~~~~~~~~ }\\
\supervisorinfo
\end{center}

\vspace{15mm}
\begin{center}
\underline{~~~~~~~~~~~~~~~~~~~~~~~~~~~~~~~~~~~~~~~~~~~~~~~~~~~~~~~~~~~~~~~~~~~~~~~~~~~ }\\
\depheadinfo
\end{center}

\vspace{5mm}
\begin{center}
\vspace{5mm}
\signdate
\end{center}

\newpage
\thispagestyle{empty}
\null\vskip0.5in
\begin{center}

  \vspace{20mm}

  \begin{LARGE}
    \dedicate
  \end{LARGE}

  \vspace{4mm}

\end{center}

\vfill

\newpage
\addcontentsline{toc}{chapter}{Acknowledgments}
\chapter*{Acknowledgments}

I am grateful to my supervisor, Professor Amir Kafshdar Goharshady, for his invaluable contributions. It was he who taught me the importance of pursuing meaningful research that in the end led to works I respect and care about. I have gained valuable insights on how to research effectively, avoid common pitfalls in academia, and plan for my future career. Also, as I began my studies, I encountered numerous obstacles which I would not have been able to overcome without the guidance and patience of Amir. Finally, I must say, I have thoroughly enjoyed our discussions throughout the past two years, particularly the ones that pushed me beyond my comfort.

I had the privilege of working and learning alongside my co-authors Zhuo Cai, S. Hitarth, Togzhan Barakbayeva, and Sergei Novozhilov for countless hours. I am grateful for their contributions, as well as the support from other colleagues in the ALPACAS research group such as Ahmed Zaher, Kerim Kochekov, Giovanna Kobus Conrado, and Pavel Hudec.

I am grateful to individuals who have played a major role in developing my passion for computers. Hossein Boomari and Mahdi Kazemi were the ones who initiated me into the world of computer science and competitive programming during my high school years. I received invaluable guidance from Professors Alimohammad Zareh Bidoki and Faezeh Ensan throughout my time as an undergraduate student. Moreover, I want to thank Behnam Shakibafar, who gave me my first job in the industry and made my experience one to cherish.

I would like to take this opportunity to express my sincere gratitude to my parents, Marzieh Javame Ghazvini and Saeid Farokhnia, who have always been there for me with their unwavering support and encouragement. I feel fortunate to have a fantastic brother, Sepehr, in my life. Even though we disagreed at times, I always have faith that he will be there for me.

As I continue my journey, I am grateful for the love and encouragement of my friends: Sanaz Safaei, Siroos Khosravinia, Hamid Sajjadi and Shima Khosravi, Mahdi Abolfazli, Hooman Khosravi, Mona Hodaei, and Kasra Pahlevani.

Finally, I am thankful to Professors Cunsheng Ding and Jiasi Shen who kindly agreed to be on my MPhil defense committee.

\newpage
\addcontentsline{toc}{chapter}{Table of Contents}
\tableofcontents

\newpage
\addcontentsline{toc}{chapter}{List of Figures}
\listoffigures


\newpage
\addcontentsline{toc}{chapter}{Abstract}
\begin{center}
{\Large \thesistitle}\\
\vspace{20mm}
by \thesisauthor\\
\departmentname\\
The Hong Kong University of Science and Technology
\end{center}
\vspace{8mm}
Smart contracts are programs that are executed on the blockchain and can hold, manage and transfer assets in the form of cryptocurrencies. Any party can interact with a smart contract by creating blockchain transactions that call the desired functions of the contract. The contract's execution is then performed \emph{on-chain} and is subject to consensus, i.e.~every node on the blockchain network has to run the function calls and keep track of their side-effects, including updates to the variables in the contract's storage, as well as transfers of cryptocurrency in and out of the contract. In most programmable blockchains, such as Ethereum, the notion of \emph{gas} is introduced to prevent DoS attacks by malicious parties who might try to slow down the network by performing time-consuming and resource-heavy computations. A fixed gas cost is assigned to every atomic operation and the originator of a function call has to pay the total gas cost of all its operations as a transaction fee.
While the gas model has largely succeeded in its goal of avoiding DoS attacks, the resulting fees are extremely high. For example, in 2022, on Ethereum alone, there has been a total gas usage of 1.77~Million~ETH $\approx$ 4.3~Billion~USD.

This thesis proposes ``lazy contracts'' as a solution to alleviate these costs. Our solution moves most of the computation off-chain, ensuring that each function call incurs only a tiny amount of gas usage, while preserving enough data on-chain to guarantee an implicit consensus about the state of the contract variables and ownership of funds. A complete on-chain execution of the functions will only be triggered in case two parties to the contract are in disagreement about the current state, which in turn can only happen if at least one party is dishonest. In such cases, our protocol can identify the dishonest party and penalize them by having them pay for the entire gas usage. In short, using our protocol, no rational party has an incentive to act dishonestly and only dishonest parties have to pay the full gas cost of the contract. If all parties are rational and therefore honest, the total gas usage of the contract shrinks dramatically.  

Additionally, we provide an implementation of our protocol as a wrapper that can be applied to any smart contract before its deployment on the blockchain. Although we focus on Ethereum contracts written in Solidity, the same technique can is also applicable to any other programmable blockchain. Finally, we perform extensive experiments over \todo{160,735} real-world Solidity contracts that were involved in \todo{9,055,492} transactions in \todo{January 2022--January 2023} on Ethereum and show that our approach reduces the overall gas usage by \todo{55.4\%}, which amounts to an astounding saving of \todo{\textbf{109.9 Million USD}} in gas fees.
\par
\noindent

\newpage
\pagenumbering{arabic}
\pagestyle{plain}
\setcounter{page}{1}
\chapter{Introduction}

\paragraph{Cryptocurrencies} Modern cryptocurrencies, such as Bitcoin and Ethereum, are decentralized monetary systems in the sense that they lack any issuing bank, government or centralized authority with control over the system and privileges beyond those of any other participant. Instead, they rely on a network of nodes that each keep track of every transaction in the system and/or the balance of every account. In such a system, all nodes should reach consensus about the composition and order of transactions and, as a result, the ownership of the funds. A user then owns a coin if and only if they own it according to the reached consensus.

\begin{figure}
	\begin{center}
	\includegraphics[keepaspectratio,width=.85\linewidth]{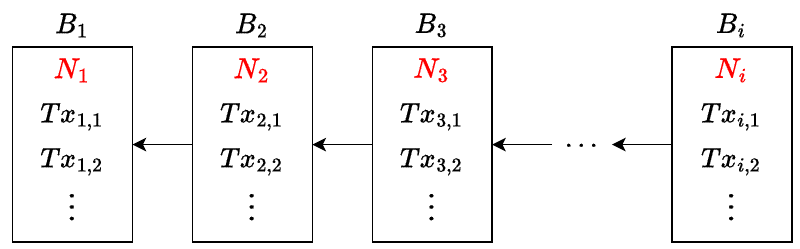}
	\end{center}
	\caption{A Simplified Representation of a Blockchain.}
	\label{fig:block}
\end{figure}

\paragraph{Blockchain} Bitcoin~\cite{nakamoto} was the first modern decentralized cryptocurrency and it achieves consensus using a proof-of-work blockchain protocol. Put simply, the transactions are grouped and ordered into blocks of a fixed maximum size. These blocks are in turn organized in a linked list using hash pointers, in which each block $B_i$ points to its previous block $B_{i-1}$ by including its hash $hash(B_{i-1}).$ This linked list of blocks is called a \emph{blockchain}. Figure~\ref{fig:block} shows a blockchain in which each block $B_i$ contains a nonce $N_i$, a pointer to the previous block $B_{i-1}$ and a sequence of transactions $Tx_{i,j}$. Any node on the network can create new transactions and blocks and publish them to the whole network using a gossip protocol. All other nodes of the network validate the transaction or block before further publishing it. For a transaction to be valid, it should spend only previously unspent funds and include valid signatures.

\paragraph{Valid Blocks and Consensus} Since we need the whole network to reach a consensus about the current blockchain, we cannot allow anyone to simply add new blocks at will, as that would fork the blockchain into many different valid chains. Bitcoin's solution to this problem is to require a proof of work, i.e. in order for a block $B_i$ to be valid, it should (a)~contain only valid transactions, and (b)~contain a nonce value $N_i$ such that the hash of the block, including its nonce field, is smaller than a specific threshold. Some nodes on the network attempt to create or \emph{mine} new blocks by continuously trying different values for $N_i$ until they find a valid block. These nodes are called \emph{miners}. To incentivize the miners to perform such long and energy-intensive operations, the protocol pays a fixed reward per successfully-mined block and the users can also include \emph{transaction fees} in each transaction that would be paid to the miner that adds the transaction to the blockchain. Finally, the consensus chain can only include valid blocks and in case there are temporarily more than one possible chain of valid blocks, then the longest one is the consensus chain. Given the high energy usage and environmental impact of proof-of-work mining, various alternatives have been proposed~\cite{DBLP:conf/fc/ParkKFGAP18,DBLP:conf/crypto/KiayiasRDO17,DBLP:conf/goodit/ToulemondeBGP22}. While some of these alternatives remove block rewards, they all continue to use transaction fees.

\paragraph{Beyond Money Transfers} Note that the blockchain protocol above can be used not only for transferring money, but also for reaching consensus about any other well-defined and deterministic computational result. Specifically, Bitcoin~\cite{nakamoto} allows scripts that can be used to set complicated conditions beyond just signature checks for accessing and spending a coin. Some protocols, notably Ethereum~\cite{wood2014ethereum}, go even further and allow arbitrary programs in a Turing-complete language.

\paragraph{Smart Contracts} In programmable blockchains such as Ethereum, a \emph{smart contract} is a program that is stored on the blockchain. See Figure~\ref{fig:excontract} as an example\footnote{A detailed description of this contract's functionality is provided in Chapter~\ref{ch:prelim}.}. Smart contracts can receive, hold and transfer money. They currently hold assets worth hundreds of billions of dollars~\cite{DBLP:conf/esop/ChatterjeeGV18}. Smart contracts allow users to call any function and the whole network keeps track of the contract's state and the money transfers incurred by such calls.

\begin{figure}
\begin{lstlisting}[language=Solidity]
contract Example {
	
	address owner = 0x5B38...beddC4;
	bytes32 desiredResult;
	
	function start(bytes32 a) public payable {
		require(msg.sender==owner);
		desiredResult=a;
	}

	function getReward(uint y) public {
		uint x = block.number;
		address payable recipient = payable(msg.sender);
		if(keccak256(abi.encodePacked(x+y))==desiredResult)
		    recipient.transfer(address(this).balance);
	}

	function cancel() public {
		require(msg.sender==owner);
		payable(msg.sender).transfer(address(this).balance);
	}

}
\end{lstlisting}
\caption{A Smart Contract written in Solidity. }

\label{fig:excontract}
\end{figure}

\paragraph{Transactions~\cite{wood2014ethereum}} To enable smart contract functionality, Ethereum-like blockchains support a wider notion of transactions than Bitcoin. More explicitly, a transaction is no longer limited to transferring money between people, but it can also: 
\begin{enumerate}[(1)]
	\item Deploy a contract's code on the blockchain, so that everyone knows about the code and the code is then immutable; or
	\item Call a function of an already-deployed contract, providing its parameters.
\end{enumerate}
See~\cite{DBLP:conf/blockchain2/Goharshady21} for a more detailed treatment and~\cite{DBLP:conf/ithings/GoharshadyBC18} for examples. Moreover, each contract has its own address that can be used for sending money, in the form of the base cryptocurrency, e.g. Ether, to it. One can also send money to a contract at the same time as calling one of its functions.

\paragraph{Execution of Function Calls~\cite{dameron2018beigepaper}} As mentioned above, all nodes in the network keep track of the blockchain, which includes the specific order of all transactions. Therefore, it is easy to reach consensus about the state of variables in every contract. Each node should just run the functions whose code is provided by transactions of type (1) with the parameters provided in transactions of type (2), respecting the order of transactions in the blockchain.

\paragraph{On-chain Computation} We say that a computation is performed \emph{on-chain} if it has to be executed by all the nodes. Computations of type (1) and (2) above are always on-chain. 

\paragraph{Gas~\cite{dameron2018beigepaper}} Given that every function call has to be executed by every node of the network in order to reach a consensus about the state of the contracts and the balances of each person/contract, the whole system is vulnerable to a denial-of-service (DoS) attack by a malicious user who creates and calls a function with large or infinite runtime. For example, such a user can deploy and call the function shown in Figure~\ref{fig:spin}. To defend against such attacks, each basic atomic operation is assigned a specific amount of \emph{gas}, roughly proportionate to its real-world cost of execution for the nodes, and the originator of each function call has to pay a transaction fee covering the overall gas usage of the on-chain computations of the call. This fee is paid to the miner, not to every node. This is so that the miners are incentivized to solve the proof-of-work problem or its variants~\cite{DBLP:conf/sac/ChatterjeeGP19}. Indeed, the miners aim to maximize their transaction fee payoffs~\cite{DBLP:conf/blockchain2/MeybodiGHS22}. Note that this fee is paid to the miner, not to every node, and is part of the transaction fee.

\begin{figure}
	\begin{lstlisting}[language=Solidity]
function spin() public {
	uint i=0;
	while(i<100000000000000000000)
		i++;
}
	\end{lstlisting}
\caption{A Solidity Function with an Intractable Runtime.}
\label{fig:spin}
\end{figure}

\paragraph{Costs of Gas} Although using gas has been successful in deterring DoS attacks, it has had the unfortunate unintended consequence of costing the blockchain users a huge amount of money in transaction fees~\cite{chen2017under} and has also been a source for many vulnerabilities~\cite{DBLP:conf/esop/ChatterjeeGV18,DBLP:conf/concur/ChatterjeeGIV18,DBLP:conf/post/AtzeiBC17}. For example, on Ethereum, which is currently the largest programmable blockchain by market capitalization~\cite{coinmarketcap}, the users paid an average of \todo{4,838.47 ETH $\approx$ 11,776,497 USD} per day\footnote{Throughout the thesis, we use exchange rates at the time of each transaction to convert its value/fee to USD. Past exchange rates were taken from Etherscan~\cite{etherscan}.} in gas costs in \todo{2022}. The Ethereum Foundation admits the problem of high gas fees in its official documentation~\cite{gas}. Due to these prohibitive costs, many smart contracts are constrained to simple programs with limited functionality in which every function terminates in a constant number of steps. There are also many protocols, including layer-two protocols, that sacrifice decentralization or trustlessness in order to reduce the gas costs. See Chapter~\ref{ch:related} for related works.

\paragraph{Layer-1 vs Layer-2 Solutions} Apart from high gas costs, there are several major scalability issues in blockchains, including low transaction rates and high latency. The main class of solutions for these problems are called Layer-1 scalability protocols and have been introduced to solve such limitations by changing the fundamental structure of the blockchain, i.e. the consensus protocol governing the main chain. In contrast, a completely distinct class of protocols, called Layer-2 protocols, build another (virtual) blockchain on top of the main blockchain to improve transaction processing rates, latency, and fees by minimizing the use of the underlying slow blockchains that are expensive and inefficient. Most transactions take place on the layer-2 chain and there are only a few that are dispatched to the main chain, meaning that the main chain only serves as an instrument for the establishment of trust among Layer-2 participants.

\paragraph{Our Contribution} In this thesis, we present a secure and trustless solution that moves most of the computations in a smart contract off-chain and ensures small gas costs for each and every function call. Moreover, it saves enough implicit data on the blockchain to be able to reconstruct the execution and final state of the contract. Simply stated, our idea is to perform the computations eagerly off-chain, but to be lazy on-chain, thus incurring less gas usage. Our method has the following advantages:
\begin{itemize}
	\item Based on our experimental results on \todo{160,735} real-world smart contracts and \todo{9,055,492} real-world transactions during \todo{January 2022--January 2023} on the Ethereum blockchain, our approach reduces the gas usage by a significant margin of \todo{\textbf{55.4 percent}}, which is equivalent to more than \todo{\textbf{109.9 Million USD}}. See Chapter~\ref{sec:exper} for experimental details.
	\item Assuming that all contract parties are honest, they are guaranteed to reach a consensus about the state of the contract at the end of its implicit off-chain execution, and the protocol will succeed without any extra costly on-chain computations.
	\item If a party or parties are dishonest, then they might not reach a consensus about the final balances and payments. Only in such cases, our protocol simulates the run of the contract on-chain. Thus, our protocol is lazy and does not perform on-chain computations unless it is forced to do so. Moreover, the on-chain simulation ensures the dishonest party is always identified. The gas is then charged to this party. Therefore, in our protocol, we have the following desirable properties: (a)~all parties are strictly incentivized to be honest\footnote{More formally, the only Quasi-strong Nash equilibrium~\cite{DBLP:conf/icbc2/ChatterjeeGP19} is when every participant is honest.}, and (b)~all dishonest behavior is identified and only the dishonest parties are penalized by having to pay high gas costs. Hence, our protocol is secure and completely risk-free for an honest participant.
	\item Our protocol is in principle applicable to virtually all smart contracts on any programmable blockchain, regardless of the language used to program them. We provide an open-source implementation of our protocol that converts a Solidity contract to a lazy contract enabling off-chain execution. This conversion process is entirely automated in a push-button tool and has no extra burden for the smart contract programmer.
	\item Our protocol is entirely trustless. We do not assume that any party can be trusted to act honestly, even though dishonest actions are penalized. It is also decentralized and has no central authority. All users have the same access and privileges and no one is at an advantage in comparison with anyone else.  
\end{itemize}

Finally, it is noteworthy that our approach is \emph{not} a layer-2 protocol. There is no layer-2 blockchain and no separate mining or consensus mechanism other than what is natively provided in layer-1.

\chapter{Background and Preliminaries}
\label{ch:prelim}

In this chapter, we provide an overview of the gas model and other relevant details of smart contracts in Ethereum. Some of these details might seem insignificant, but handling them correctly is crucial to ensuring that our protocol in Chapter~\ref{ch:method} is sound and faithfully models/simulates the original contract. 

\paragraph{Languages} Ethereum smart contracts are stored on the blockchain in a stack-based assembly-like format called EVM (Ethereum Virtual Machine) bytecode~\cite{dameron2018beigepaper}. Various high-level languages are compiled to this bytecode, such as Solidity~\cite{solidity}, which is a strongly-typed language loosely inspired by Javascript, and Vyper~\cite{vyper}, which is similar to Python. Solidity is currently the most widely-used language for writing smart contracts in Ethereum~\cite{langs, DBLP:conf/sac/ChatterjeeGG19}. Figure~\ref{fig:excontract} shows an example contract written in this language. In this work, we focus our presentation on contracts written in Solidity since they are more human readable than the bytecode. Nevertheless, every step of our approach is in principle applicable to EVM bytecode, Vyper, and any other language on other programmable blockchains, too. 

\paragraph{Functions~\cite{solidity}} Each contract consists of a number of functions. For example, Figure~\ref{fig:excontract} shows a contract with three functions. A function can be public, meaning that it can be invoked by anyone who creates a transaction calling this function, as well as other smart contracts, or it can be private to the contract, i.e.~only callable by other functions in the same contract. One can also limit a function to be callable only by other contracts or by derived contracts.

\paragraph{Memory and Storage~\cite{solidity}} There are two types of space for storing data in a contract: (i)~a \emph{memory} which is the working space of the contract and is erased after each transaction, and (ii)~a persistent \emph{storage} whose contents are not erased in between transactions and are permanently stored by all nodes on the network. On some blockchains, the types of memory are further refined. However, for the purposes of this work, we do not need to consider a finer division. 

\paragraph{Types of Variables~\cite{solidity}} Accordingly, there are three types of variables in a Solidity smart contract:
\begin{itemize}
	\item \emph{State variables} are the ones saved in the storage. The values of these variables collectively define the \emph{state} of the contract. For example, in Figure~\ref{fig:excontract}, \texttt{owner} and \texttt{desiredResult} are state variables. Every node on the blockchain keeps track of the values of these variables at all times.
	\item \emph{Local variables} are allocated in the memory and used only during a single execution. Their values are discarded after the transaction comes to an end. For example, in Figure~\ref{fig:excontract}, \texttt{x} is a local variable in \texttt{getReward}. 
	\item \emph{Global variables} are special variables that provide information about the current block and the state of the blockchain. For example, \gv{block.number}, which is used in the \texttt{getReward} function of Figure~\ref{fig:excontract}, provides access to the index of the current block, i.e.~the block containing the current transaction. Similarly, \gv{msg.sender} provides the address of the caller of the current function. See~\cite{global,dameron2018beigepaper} for a list of global variables.
\end{itemize}

\paragraph{Example} We are now ready to illustrate the contract in Figure~\ref{fig:excontract} in more detail. This contract has an owner whose address is hard-coded as a state variable. The owner can start a competition by calling the function \texttt{start} and providing a value \texttt{a} which will in turn be saved in the state variable \texttt{desiredResult}. This function is tagged as \keyword{payable}, which means the caller can include a monetary payment while calling this function and the money would become the property of the contract. The function also checks that its caller, \gv{msg.sender}, is the owner of the contract. Otherwise, the \keyword{require} statement fails, throwing an exception which will cause the current transaction and all its side-effects to be reverted. In other words, only the owner can call this function. The competition's goal is to find an integer \texttt{y} such that when it is added to the current block number and hashed using the \texttt{keccak256} hash function, it leads to the \texttt{desiredResult}. Anyone who finds such a value can call \texttt{getReward}. This function checks the hash and if it has the desired value, then transfers the entire balance of the contract to the caller. Additionally, the owner can call the \texttt{cancel} function to cancel the competition and withdraw the contract's balance.

\paragraph{Ethereum's Gas Model~\cite{wood2014ethereum}} On Ethereum, executing each EVM bytecode operation costs a well-defined number of units of \emph{gas}. See~\cite[Appendix A]{dameron2018beigepaper} for a complete table of gas costs. Generally, operations on storage are much costlier than memory. When a user initiates a transaction, she can set two values: (a)~the maximum amount $g_m$ of gas that she is willing to pay for, and (b)~the price $\pi$, in ether, that she is willing to pay for each unit of gas. Based on these, a miner can choose which transactions to include in her block and in which order. When the transaction is included in a block and mined, every node on the blockchain executes it. This is called on-chain execution and begins by taking a deposit of $g_m \cdot \pi$ from the initiator's account and then running the called function, while keeping track of the total gas used until this point. There are three cases:
\begin{enumerate}
\item If the function terminates using $g$ units of gas and $g \leq g_m,$ then the miner is paid a transaction fee of $g \cdot \pi$ and the rest of the deposit, i.e.~$(g_m-g)\cdot \pi$, is reimbursed to the initiating user. Moreover, any updates to the storage are saved by all nodes on the blockchain;

\item If the function throws an exception/error, then all of its effects are reversed and the storage is returned to its status before this function call and the gas deposit is divided exactly as in the last case; 
\item Otherwise, an out-of-gas exception is triggered, causing the user to lose her deposit, which is paid to the miner in its entirety, and the storage state of all affected contracts to revert to right before the current transaction. Out-of-gas errors are a common and serious security vulnerability in smart contracts~\cite{atzei2017survey,DBLP:journals/corr/abs-2112-14771,DBLP:conf/tacas/AlbertCGRR20,DBLP:journals/access/AshrafMJC20}. However, avoiding them is an orthogonal issue and our approach ensures the exact same behavior in the optimized low-gas version of the contract as in the original.
\end{enumerate}

In all the three cases above, there will be a network-wide consensus about the balance of every contract and user, including the miner, as well as the storage of all contracts. This is precisely due to on-chain execution, i.e.~the fact that every node has executed the function calls and kept track of their effects and gas usage as described above.

\paragraph{London Upgrade~\cite{gas}} Since August 2021, the users have to pay an additional \emph{base fee} for their transactions and part of the transaction fees are burned and not paid to the miner. However, these changes do not affect our protocol and can be safely ignored in the sequel.

\paragraph{Gas Sharing} A function in a smart contract can itself call another function, either in the same contract or in a different contract. In such cases, the caller function can decide how much of its gas it wants to allocate to the callee, but it cannot change the gas price $p$. Moreover, it can decide what happens if the callee throws an exception and is reverted, i.e.~the caller might also throw the same exception and get reverted or it might catch the exception and continue its own execution. See~\cite{solidity} for details.

\paragraph{Block Size~\cite{dameron2018beigepaper}} Ethereum blocks have a maximum amount of gas they can use, i.e. the sum of all transactions in a single block. The limit can change based on the demand in the network but is at most 30 million units of gas per block. 
\chapter{Our Protocol: Lazy Contracts}
\label{ch:method}

In this chapter, we present our protocol to move most of the computations off-chain and significantly reduce the gas usage of a given smart contract.
From now on, we assume that a smart contract $\contract$ is given and that it is written in the Solidity language. We start by illustrating the fundamental idea and then follow up with the technical details of the protocol.

\section*{Fundamental Idea}
The main idea behind our approach is quite simple and elegant. Let $\parties$ be the set of parties who intend to interact with the contract $\contract.$ Then, it suffices to ensure that all members of $\parties$ are in agreement about the current state of $\contract$ and there is no need to force every other node in the network to constantly keep track of $\contract$'s state, as well. Such an agreement in $\parties$ can be obtained off-chain by storing only a small amount of information on-chain. More specifically, if we store $\contract$'s code and a list of all function calls on-chain, without actually executing the function calls on-chain, then (i)~we avoid paying gas fees for the on-chain execution of function calls, (ii)~any member of $\parties$ can execute all the function calls in the right order off-chain, i.e.~on her own machine, and they will all reach the same final state, and (iii)~this final state is uniquely determined by the information that is stored on the blockchain. So, if a party $\alice \in \parties$ is dishonest, we can run all the function calls on-chain, identify the dishonest party $\alice$, and penalize her by billing her the entire gas cost. In short, instead of actually executing everything on-chain, we can just store the function calls' information, i.e.~which function was called and with what parameters, without executing the call.

\subsection*{Lazy Execution of Function Calls}
Our idea above is a variant of the well-known concept of lazy evaluation~\cite{DBLP:journals/csur/Hudak89,DBLP:conf/popl/Launchbury93,DBLP:conf/pldi/Johnsson84,DBLP:phd/ndltd/Maessen02}. However, in the context of smart contracts, we take lazy evaluation to the extreme and apply it (i)~only to on-chain computations, and (ii)~at the level of function calls. This is due to the fact that we have a clear distinction between on-chain and off-chain computation. Off-chain computation does not incur gas costs and is performed only by the parties to the contract on their own machines. Since it is cheap, we perform it eagerly, i.e.~each party $\paul \in \parties$ executes a function call as soon as possible and updates his own copy of $\contract$'s state. In contrast, on-chain computation is performed by the miners and all the other nodes on the network and thus costs significant gas. As such, it is expensive and should be performed only when necessary. Hence, we opt for a lazy approach that triggers an on-chain execution only when two parties $\alice, \bob \in \parties$ have a disagreement about the current state. Such a disagreement is only possible if at least one side is dishonest. When we perform on-chain computations, we will always identify the dishonest party, who will then bear all the gas costs. 

\subsection*{Game-theoretic Guarantee}
Based on above, it is clear that a dishonest party is heavily penalized by having to pay for the gas that is used in the on-chain execution of all function calls, including function calls by other parties. As such, there is no incentive to act dishonestly and no rational party would do so. In practice, this guarantees that the on-chain execution of function calls is unnecessary in most cases and happens only if one party is willingly and actively trying to lose money. Even in such a strange and unlikely case, the on-chain computation will not have any adverse financial effect on the other parties as the dishonest party bears all its costs.

\section*{Our Protocol}
Our protocol is implemented as a wrapper ``lazy contract''
 $\wrapped,$ also in Solidity, which includes a slightly modified version of the code of $\contract,$ as well as additional functionality. The developer should deploy $\wrapped$ to the blockchain, instead of directly deploying $\contract.$ Obtaining $\wrapped$ from $\contract$ is a well-defined algorithmic process and we provide a push-button tool that performs this task automatically (See Chapter~\ref{sec:exper}). More specifically, the contract $\wrapped$ has the same state variables as in $\contract,$ several new state variables for the new functionality, and for every function $\contract.f$ there is a corresponding function $\wrapped.f$ with minor changes, described further below. Additionally, $\wrapped$ allows the developer to set values for the following state variables upon its deployment on the blockchain:
\begin{itemize}
	\item The deposit $d$ that each user should put down to ensure the gas costs of on-chain execution can be billed to this user in case of dishonest behavior;
	\item A positive integer $t$ used as a time limit for challenges, whose use-case will become apparent in the withdrawal and challenge process described below;
	\item The maximum amount of gas that each user is allowed to consume in executing functions of $\contract$;
	\item The maximum amount of gas that each single function call to a function in $\contract$ is allowed to consume;
	\item The maximum number of allowed function calls of $\contract$.
\end{itemize}
The developer can choose not to enforce one or more of the maximum values above. However, (i)~the limits must ensure that each function call can be executed within one block (cf. ``block size'' in Chapter~\ref{ch:prelim}), and (ii)~it is the developer's responsibility to ensure that the deposit $d$ is enough to cover all potential gas usage in the contract's lifetime. Using the values chosen by the developer, $\wrapped$ provides the following additional functionality:
\subsection*{Joining}
Before being able to interact with $\contract$'s functionality in $\wrapped$, a user/party $\paul \in \parties$ has to explicitly join $\wrapped$ by calling $\wrapped.\texttt{join()}$ and providing a deposit of $d$ ether. This deposit will remain in the contract's custody as long as $\paul$ is a party to the contract and will be used to compensate for gas usage if $\paul$'s dishonest behavior triggers an on-chain execution. Note that joining is an on-chain action and $\paul$ pays for its gas. However, the gas usage is bounded by a small constant since this is a one-off function call with $O(1)$ runtime and gas usage.
\subsection*{Virtual Banking}
Since $\wrapped$ is lazy in running calls to functions of $\contract$ and avoids running them on-chain by default, any transfer of money by these function calls is also not executed on-chain. To enable the functionality of money transfers between $\contract$ and the participants, $\wrapped$ acts as a bank and allows each joined user to deposit and withdraw ether to $\wrapped.$ The balances used in $\contract$'s functions will then refer to the user's balance in $\wrapped,$ rather than her ether balance on the blockchain. These $\wrapped$-balances are not explicitly stored on-chain. Each party in $\parties$ keeps track of them off-chain on their own machine. The contract $\contract$ has a $\wrapped$-balance as well.

\subsection*{Ledger}
$\wrapped$ keeps track of an on-chain internal ledger (as a state variable) which is a sequence of deposits and withdrawals by the users to $\wrapped$'s bank and also the function calls requested by the users to functions of $\contract.$ 

\subsection*{Depositing Ether} 
A party $\paul \in \parties$ can deposit ether to $\wrapped$ at any time. The deposited ether will be under the control of $\wrapped$ and an entry will be added to $\wrapped$'s ledger certifying the amount that $\paul$ deposited. This entry is added on-chain. Upon seeing this entry, all participants in $\parties$ update their off-chain version of \paul's $\wrapped$-balance accordingly.
\subsection*{Lazy Function Call}
$\wrapped$ has a dedicated function which is named \texttt{requestCall} and can be utilized by a party $\paul$ who wants to call a function in $\contract.$ To call $\contract.f,$ $\paul$ has to create a transaction that calls $\wrapped.\texttt{requestCall}$ and provides the following parameters:
	\begin{enumerate}[a.]
		\item The name $f$ of the function that should be called,
		\item The maximum amount $g_m$ of gas that may be used by~$f$,
		\item The parameters that should be passed to $f,$
		\item the amount of money that should be paid from $\paul$'s $\wrapped$-balance to $\contract$'s $\wrapped$-balance. This is only applicable if $\contract.f$ is payable, i.e.~if it can accept payments.
	\end{enumerate}
	Upon receiving the items above, $\wrapped.\texttt{requestCall}$ first checks that $\paul$ is not exceeding the maximum gas usage allowed by the developer. If this limit is exceeded, then the function call is ignored. Otherwise, instead of running $\contract.f$ or $\wrapped.f$ on-chain, \texttt{requestCall} adds an entry to the internal ledger. This entry includes all the parameters (a-d) above, as well as a record of the values of all global variables\footnote{In practice, our tool performs a lightweight syntactic analysis and ensures that $\wrapped$ only saves the values of those global variables that are used in $f.$\medskip} such as \gv{block.number} and $\gv{msg.sender} = \paul$. This is all the information that can possibly be needed for executing the call to $f,$ but the call itself is \textbf{\emph{not}} executed on-chain. In other words, we are lazy when it comes to on-chain computations. 
	
	When a call request record is added to the ledger in $\wrapped$, every party in $\parties$ performs that function call off-chain on their own machine and updates their own copy of the state variables in $\contract$ and the $\wrapped$-balances accordingly. Thus, we are eager to execute the function calls off-chain. Put simply, instead of running the functions on-chain, they are simulated in a virtualized off-chain environment by each of the parties. This ensures no gas is paid for the execution of the functions and the only gas costs are the ones triggered by adding an on-chain record of the function call parameters to the ledger of $\wrapped.$\footnote{Note that, although the function call is not executed on-chain and $\paul$ does not need to pay for its gas, he still has to provide a maximum gas usage amount $g_m$. This is a technical detail with no significant practical implications, but cannot be ignored for two reasons: (i)~the semantics of Ethereum contracts depend on gas. Thus, the parties need to know $g_m$ to perform their off-chain execution; and (ii)~An on-chain execution might be triggered in the future.}
	
\begin{figure}[H]
		\begin{center}
		\includegraphics[keepaspectratio,width=0.7\linewidth]{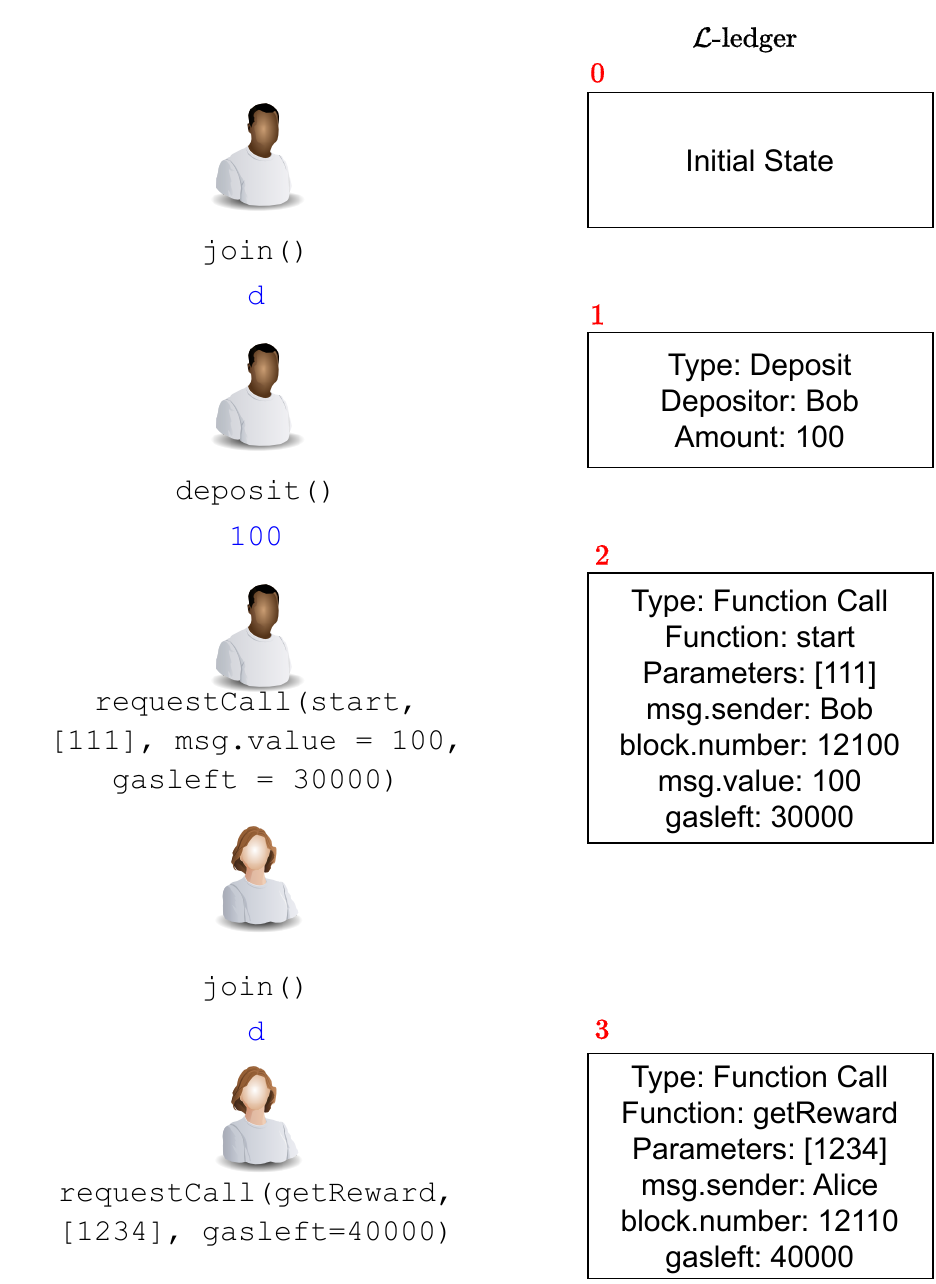}
		\end{center}
		\caption{Interactions of $\alice$ and $\bob$ with a Lazy Contract $\wrapped$ over the Contract $\contract$ of Figure~\ref{fig:excontract}. }
		\label{fig:ex1}
	\end{figure}

	\paragraph{Example}
		Consider the contract $\contract$ of Figure~\ref{fig:excontract}. The developer first obtains the wrapped lazy contract $\wrapped$ and deploys it on the blockchain. Suppose that the owner is $\bob,$ i.e.~$\bob$'s address is hard-coded in $\contract.$ Figure~\ref{fig:ex1} shows an example interaction of $\alice$ and $\bob$ with $\wrapped.$ The function calls that are executed on-chain are shown on the left and their effect on the $\wrapped$-ledger is shown on the right. Payments to $\wrapped$ are shown in blue. First, $\bob$ joins $\wrapped.$ Then, he deposits $100$ units of currency to $\wrapped,$ so that his $\wrapped$-balance is now $100.$ This is recorded at index $1$ of the $\wrapped$-ledger. He then requests the execution of the function $\contract.\texttt{start}(111),$ i.e.~providing a value of $111$ for the parameter $a$. Since $\contract.\texttt{start}$ is payable, Bob is also requesting that $100$ units of his $\wrapped$-balance be paid to $\contract$ at the time of this function call. This is the {\gv{msg.value}}. Similarly, he is setting a maximum gas usage of $30000$ for any future simulation of this function call. All these parameters are recorded in the $\wrapped$-ledger, together with current values of all other global variables, e.g.~\gv{block.number}. However, no call is made to the $\texttt{start}$ function itself. Instead, a record is added to the $\wrapped$-ledger that includes all the necessary data for performing such a call in the future, if needed. Then, $\alice$ joins $\wrapped$ and similarly requests a call to the function $\texttt{getReward}.$
	
\subsection*{Withdrawing Ether}
The parties can decide to withdraw ether from their $\wrapped$-balance at any time so that they can use it elsewhere on the blockchain. However, this is a bit tricky since the $\wrapped$-balances are not explicitly stored in $\wrapped$ and are dependent on the function calls that were lazily added to the ledger instead of being executed on-chain. Nevertheless, all other parties have executed these function calls off-chain and are hence aware of each other's $\wrapped$-balances. Thus, we adopt a two-step process:
\begin{itemize}
	\item Step 1: The party $\alice \in \parties$ calls  $\wrapped.\texttt{requestWithdraw}(x)$ and specifies the amount $x$ she wishes to withdraw as a parameter. This request is added to the $\wrapped$-ledger.
	\item Step 2: If no other party challenges the withdrawal until $t$ blocks after Step 1, then $\alice$ can call the dedicated function $\wrapped.\texttt{withdraw}$ and receive the desired amount of money. Recall that $t$ was a parameter set by the developer.
\end{itemize}

\begin{figure}[H]
	\begin{center}
	\includegraphics[keepaspectratio,width=0.7\linewidth]{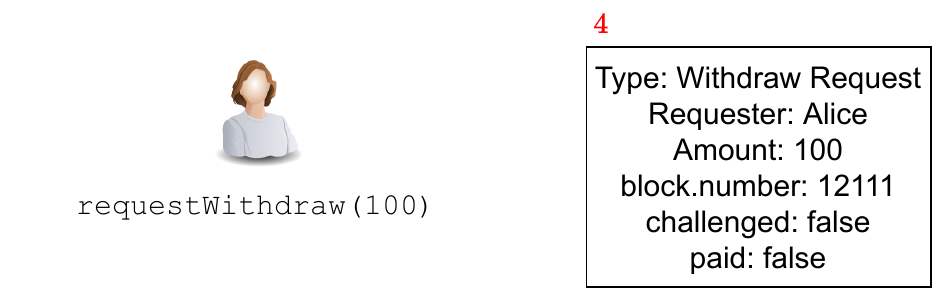}
	\end{center}
	\caption{$\alice$ Requests to Withdraw Money.}
	\label{fig:ex2}
\end{figure}

\paragraph{Example}
	Continuing with Figure~\ref{fig:ex1}, suppose that $\alice$ requests to withdraw $100$ units. This is shown in Figure~\ref{fig:ex2}. This request will be added to the $\wrapped$-ledger, together with the number of the blockchain block in which the request was made and two boolean variables keeping track of whether the request is challenged and whether it is already paid out. This request was made at block 12111. Suppose the developer has set $t=100.$ So, any other party, e.g.~$\bob$, can challenge this request until block 12211. This would set the \emph{challenged} boolean value to true. Anytime after block 12211, $\alice$ can call $\wrapped.\texttt{withdraw}(4)$ and get her 100 units of money assuming both \texttt{challenged} and \texttt{paid} are false. When she receives the money, \texttt{paid} will be set to true to avoid double payment. Note that $\alice$ can receive the money as long as $\bob$ does not challenge her and this would not require an on-chain execution of the functions. In other words, $\bob$'s inaction in the $t$ blocks after the request would be interpreted as his approval that $\alice$'s $\wrapped$-balance was at least $100$ at the time of the request. 

\subsection*{Challenging}
Suppose a party $\alice \in \parties$ requests a withdrawal of $x$ ether from her $\wrapped$-balance. As mentioned above, this request will be added to the ledger in $\wrapped.$ However, the $\wrapped$-balances are not explicitly stored in $\wrapped$ on-chain and are instead implicit and depend on the whole sequence of operations that are recorded on the $\wrapped$-ledger. Recall that these operations (lazy function calls, deposits, withdrawals) are all eagerly simulated by every other party $\bob \in \parties$ off-chain on his own machine. Thus, $\bob$ knows whether $\alice$ has a $\wrapped$-balance of at least $x$. If $\bob$ finds out that $\alice$ is trying to withdraw more than her $\wrapped$-balance, then $\bob$ can challenge the withdrawal by calling a function named $\wrapped.\texttt{challenge}(j).$ This function takes the index $j$ of the withdrawal request by $\alice$ in the $\wrapped$-ledger.

When a challenge occurs, then either $\alice$ is dishonest and trying to withdraw more than her balance or $\bob$ is dishonest and stopping $\alice$ from withdrawing her money. In such a case, the lazy contract $\wrapped$ initiates an on-chain evaluation (discussed in more detail below). The on-chain evaluation will run all the operations listed in the $\wrapped$-ledger on-chain, potentially causing a huge gas cost. However, this ensures that the $\wrapped$-balances are also computed on-chain and hence it is easy to check whether the $\wrapped$-balance of $\alice$ was at least $x$ at the time she requested to withdraw $x$ units. This will directly show whether $\alice$ was dishonest or $\bob$. Suppose the total gas cost for the on-chain evaluation is $g \cdot \pi$. The lazy contract $\wrapped$ pays for $g \cdot \pi$ from the $d$ units of the deposit that were put down by the dishonest party when they joined the contract in (1) above. The dishonest party is blacklisted and will not be able to interact with $\wrapped$ until they make their deposit whole again by paying $g \cdot \pi$ using $\wrapped.\texttt{join()}.$
If a withdrawal request is not challenged within $t$ blocks, where $t$ is the constant chosen by the developer, the requesting party $\alice$ can withdraw her money from $\wrapped.$

\begin{figure}[H]
	\begin{center}
	\includegraphics[keepaspectratio,width=0.7\linewidth]{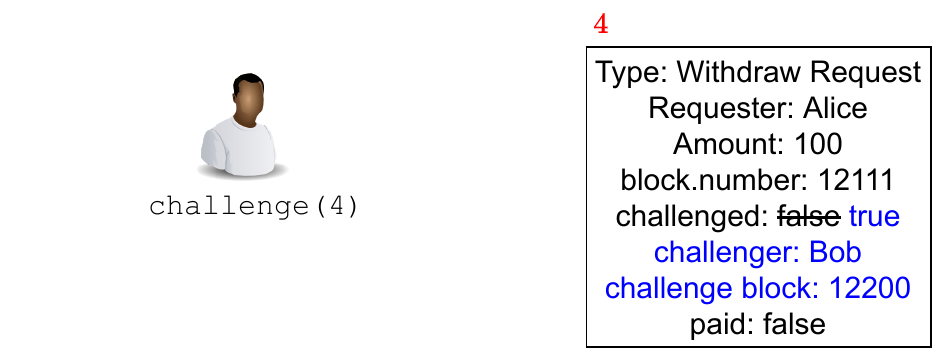}
	\end{center}
	\caption{$\bob$ Challenges $\alice$'s Request.}
	\label{fig:ex3}
\end{figure}

\paragraph{Example}
	Continuing from Figure~\ref{fig:ex2}, suppose that $\bob$ challenges $\alice$'s withdrawal. He should call $\wrapped.\texttt{challenge}(4),$ because the withdraw request was added to index 4 of the $\wrapped$-ledger. The effect of this challenge is shown in Figure~\ref{fig:ex3}. Note that the block number (time) when Bob has issued the challenge is also recorded on the ledger. Of course, this should be within $t$ blocks of the original withdrawal request.

\subsection*{Leaving}
Any party can leave the contract at any time as long as they do not have an active challenge or withdrawal request. A withdraw request is active if (i)~the current time is within $t$ blocks after the request, or (ii) the request has been challenged and the ensuing on-chain evaluation has not concluded yet. The leaving party can call $\wrapped.\texttt{leave()},$ which checks for the existence of active withdraw or challenge requests. If there are no such requests, the leaving party will get their deposit $d$ back and would not be able to interact with $\wrapped$ unless they rejoin using $\wrapped.\texttt{join()}$ and repay the deposit.
\subsection*{On-chain Evaluation}
On-chain evaluation is triggered only when there is a disagreement about the state of $\contract$ and a withdrawal is challenged. Let $i$ be the index of the last operation in the $\wrapped$-ledger that is executed on-chain. Originally, we have $i=0,$ but $i$ might have increased in case of previous challenges. Let $j$ be the index of the withdrawal request by $\alice$ in the $\wrapped$-ledger and suppose the request is challenged by $\bob.$ We have to simulate all the operations in the range $[i+1, j]$ of the $\wrapped$-ledger \emph{on-chain} in order to find the $\wrapped$-balance of $\alice$ at the time of the request. The contract $\wrapped$ maintains a state variable $b[\paul]$ for every party $\paul \in \parties.$ This is the $\wrapped$-balance of $\paul$ immediately after the on-chain execution of the $i$-th entry in the $\wrapped$-ledger, i.e.~his $\wrapped$-balance up until the last operation that is executed on-chain. It also has a state variable $b[\contract]$ which similarly tracks the $\wrapped$-balance of $\contract.$ If we execute everything until index $j-1$ on-chain, then $b[\alice]$ would be the $\wrapped$-balance of $\alice$ at the time she requested to withdraw $x$ units. So, we simply have to figure out whether $b[\alice] \geq x.$
	
Note that a smart contract cannot initiate its own execution and all function calls have to be initiated by a user. Moreover, the initiator of a function call has to pay for its gas usage. However, we would like to ensure that (i)~the dishonest party ultimately bears the gas costs\footnote{Recall that we assume the deposit $d$ is chosen such that it is enough to cover all possible gas costs.}, and (ii)~the incurred gas costs are as small as possible. Hence, $\wrapped$ holds an auction for the role of initiator in which anyone can enter a bid for the gas price $\pi$ that they are willing to charge for the execution. More specifically, the on-chain simulation contains the following steps:
\begin{enumerate}
	\item \emph{Bidding.} For $t$ blocks after the challenge, any party $\ingrid \in \parties$ can call the function $\wrapped.\texttt{bid(}j, \pi\texttt{)}$ signifying that she is willing to initiate the on-chain execution as long as she is paid $\pi$ units of currency (ether) per unit of gas. Note that this is not necessarily the amount that she really pays the miners per unit of gas. Indeed, $\pi$ can be larger than the real unit gas cost, hence giving some profit to $\ingrid$ who is volunteering to initiate the on-chain execution. However, this profit is unlikely to be high as parties can undercut each other. The potentially dishonest parties $\alice$ and $\bob$ cannot bid\footnote{However, they can join using a new account (identity) and then bid. This is not a problem, since they will have to pay another deposit $d$ in order to join $\wrapped$ with a different identity.}. After $t$ blocks, the user $\ingrid$ who made the smallest bid becomes the initiator and triggers on-chain simulation.

	\item \emph{Simulating Every Entry of the $\wrapped$-ledger.} After $\ingrid$ is chosen in the bidding process above, she has to call $\wrapped.\texttt{simulate}(k)$ for every $i+1 \leq k \leq j$ in order. This function simulates the execution of the $k$-th entry in the $\wrapped$-ledger \emph{on-chain}. This will of course incur gas costs which are paid by $\ingrid$. However, $\wrapped.\texttt{simulate}$ keeps track of how many units of gas $\ingrid$ has paid in a state variable $\gamma[\ingrid].$ If $\ingrid$ fails to call $\wrapped.\texttt{simulate}(k)$ within $t$ blocks after she called $\wrapped.\texttt{simulate}(k-1),$ she loses her deposit $d$ and cannot interact with $\wrapped$ unless she pays the deposit again\footnote{In practice, this might be too harsh a punishment and one can confiscate only part of the deposit $d$ from $\ingrid$.}. Moreover, another bidding process will begin as in Step 1 above to find a new initiator for the rest of the on-chain simulations. This ensures that the simulation will be successfully carried out until index $j$.
\end{enumerate}

\paragraph{Example}
Continuing from Figure~\ref{fig:ex3}, since $t=100$ and the challenge was at block 12200, the contract $\wrapped$ accepts bids for simulating the first four entries in the $\wrapped$-ledger from time (block number) 12200 to time 12300. The owner $\ingrid$ of the lowest bid can then initiate the simulations. She should call $\wrapped.\texttt{simulate}(1)$ successfully before time 12400 and then each $\wrapped.\texttt{simulate}(k)$ should be called within 100 blocks of $\wrapped.\texttt{simulate}(k-1)$ for every $2 \leq k \leq 4.$ Otherwise, we will have another round of bidding and a different initiator will be chosen. Let's assume that $\ingrid$ acts fast and calls $\wrapped.\texttt{simulate}(k)$ at time $12300+k$ for every $1 \leq k \leq 4.$ See Figure~\ref{fig:ex4}.

\section*{Details of the Simulation} 
$\wrapped.\texttt{simulate}(k)$ has to perform different operations depending on the type of entry at index $k$ of the $\wrapped$-ledger. The details of these operations are as follows:
\begin{itemize}
	\item \emph{Function Call.} If the $k$-th entry is a function call to $\contract.f$, then $\wrapped.\texttt{simulate}(k)$ calls $\wrapped.f$ and provides it with the parameters and global variables that were recorded by \texttt{requestCall} when this entry was first added to the ledger. Moreover, the gas usage in this call will be limited to the \gv{gasleft} parameter provided by the requestor. If $f$ is \keyword{payable}, then $\wrapped.\texttt{simulate}(k)$ checks that the caller has enough $\wrapped$-balance for the payment to $f$. Otherwise, the call is ignored.  $\wrapped.f$ will then simulate the execution of $\contract.f$ as if it were called at the time of the original \texttt{requestCall} and in that same environment. 
	\item \emph{Deposits.} If the $k$-th entry is a deposit by party $\paul$ then $\wrapped.\texttt{simulate}(k)$ simply increases $b[\paul]$.
	\item \emph{Withdraw Request.} Similarly, if the $k$-th entry is a request by a party $\paul$ to withdraw an amount $y$, then $\wrapped.\texttt{simulate}(k)$ first checks to see if $b[\paul] \geq y,$ i.e. whether the withdrawing person has enough balance. If so, it decreases $b[\paul]$ by $y$. If not, it ignores the withdraw request and lets $b[\paul]$ remain unchanged\footnote{Effectively, this is equivalent to assuming only valid withdraw requests, i.e.~requests that are made at a time that the requester has enough balance, are processed and that every other request is challenged and defeated and hence ignored. Also note that we assume the withdrawal happens at the time of request and decrease the $\wrapped$-balance immediately as the request comes in. The fact that it takes $t$ blocks for the withdrawal to be successful does not mean the money is available during those $t$ blocks. The parties should consider this when doing off-chain simulations. \medskip}.  If $k=j,$ i.e.~if this is the challenged withdrawal of $\alice$, then $\wrapped.\texttt{simulate}(j)$ checks to see if the withdrawal was successful. If so, $\alice$ (the requester) is honest and $\bob$ (the challenger) is dishonest. Otherwise, $\bob$ is dishonest and $\alice$ is honest. In any case, the gas cost $\gamma[\ingrid] \cdot \pi$ is paid to the on-chain initiator $\ingrid$ and deducted from the deposit of the dishonest person\footnote{If there are several initiators, they each get their own share of the gas costs. The amount is calculated but not paid directly. There is instead a function called $\wrapped.\texttt{getGasPayment()}$ that can be called by $\ingrid$ to receive the payment. This is to follow the pull-over-push security recommendation in Ethereum smart contracts~\cite{pullpush}.}. 
\end{itemize}

While an on-chain evaluation is in process, the other functionalities of $\wrapped$ are intact and other users/parties who are not part of the dispute can continue to use $\wrapped$ as normal.

\begin{figure}
	\begin{center}
		\includegraphics[keepaspectratio,width=0.95\linewidth]{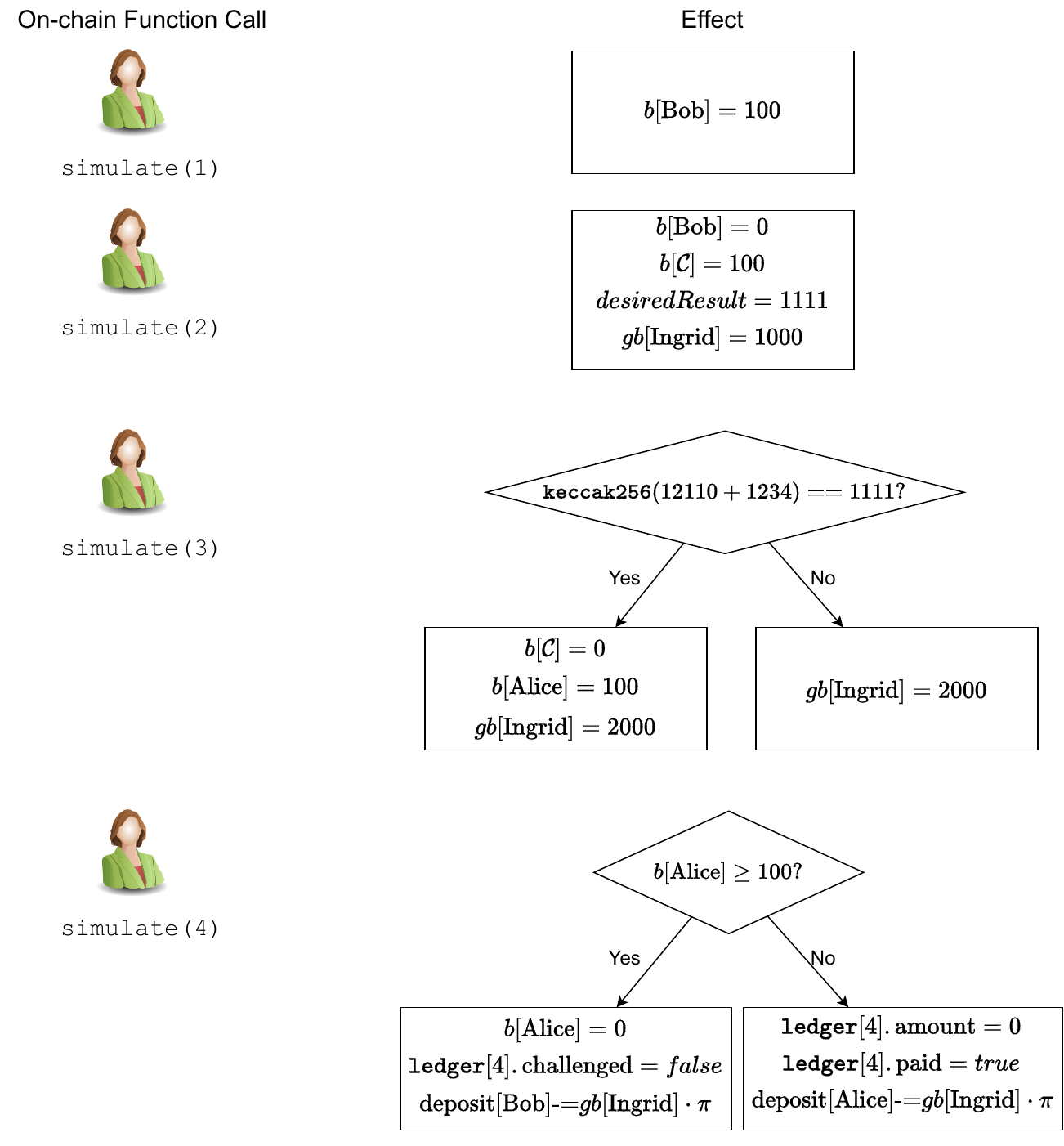}
	\end{center}
	\caption{On-chain Function Calls of an Initiator ($\ingrid$) and Their Effects.}
	\label{fig:ex4}
\end{figure}

\paragraph{Example}
	Following from previous examples (Figures~\ref{fig:ex1}--\ref{fig:ex3}), $\ingrid$ makes the function calls shown in Figure~\ref{fig:ex4}, simulating the entries in the $\wrapped$-ledger on-chain in order to establish which of $\alice$ and $\bob$ are dishonest. For simplicity, let's assume every function call uses exactly $1000$ units of gas. When simulating the first entry in the ledger, the only effect is to increase $b[\bob].$ The second entry is the function call by $\bob.$ The called function \texttt{start} is \keyword{payable} and $\bob$ is paying $100$ units to the contract $\contract$. Hence, $b[\bob]$ decreases by $100$ and $b[\contract]$ increases by $100.$ Moreover, this function call sets the variable $\texttt{desiredResult}$ to $1111$ as per the parameters provided by $\bob.$ Finally, $\gamma[\ingrid]$ is increased by the amount of gas that was used for this function call. Similarly, $\texttt{simulate}(3)$ runs the function call requested by $\alice,$ i.e.~$\contract.\texttt{getReward}(1234).$ However, $\alice$'s request was added to the ledger at time $12110,$ so this is the value used for $\texttt{x}$ in the execution of $\texttt{getReward}.$ Following this function, if the resulting hash value is correct, $\alice$'s balance increases to $100$ and the contract $\contract$'s balance decreases accordingly. Otherwise, there is no change in the balances. In any case, $\wrapped$ also keeps track of $\ingrid$'s gas money. Finally, when $\texttt{simulate}(4)$ is called, if $\alice$'s balance is enough for the withdrawal requested at index $4$ of the ledger, then $\bob$ was dishonest. In this case, $\bob$'s deposit is deducted to pay $\ingrid$'s gas money. Moreover, the withdrawal request at index $4$ is no longer challenged, so $\alice$ can withdraw her money. Otherwise, if $\alice$ did not have enough balance for the withdrawal, then she was the dishonest party. In this case, $\alice$ will be penalized and her deposit used to pay $\ingrid$'s gas money. Also, the withdrawal amount is set to $0$ and the $\texttt{paid}$ flag is set to true to ensure $\alice$ would not be able to withdraw money from $\wrapped.$

\subsection*{Changes to $\contract$'s Code in $\wrapped$} 
We mentioned that the wrapper lazy contract $\wrapped$ has a function $\wrapped.f$ corresponding to each function $\contract.f$ in the original contract $\contract.$ The function $\wrapped.f$ is quite similar to $\contract.f$ but has some important differences. First, it uses the $\wrapped$-balances instead of normal balances on the blockchain. Second, it has to mimic the behavior of $\contract.f$ at the time of $\texttt{requestCall()}$ rather than the current time, i.e.~the time of $\texttt{simulate().}$ In other words, it has to use the right values for the global variables. To handle these two cases, every expression of the form $\paul.\keyword{balance}$ in $\contract.f$ is replaced by $b[\paul]$ in $\wrapped.f.$ Recall that $b[\paul]$ is a state variable keeping track of the $\wrapped$-balance of $\paul$ until the point of the ledger that is executed on-chain. Similarly, a payable function call by $\paul$ reduces $b[\paul]$ and a \keyword{transfer} or \keyword{send} that sends $x$ units of money to $\paul$ is replaced by $b[\paul]\texttt{+=}x.$ Moreover, every global variable $\gv{v}$ is replaced by a function call $\wrapped.\texttt{get\_v}(k)$ that returns the value of $\gv{v}$ that was saved at index $k$ of the ledger, i.e.~its value at the time of the $\texttt{requestCall}.$ These changes ensure that $\wrapped.f$ simulates $\contract.f$ correctly. Note that all exception-handling behavior remains unchanged. Finally, if $f$ is a function in the original contract $\contract,$ then $\wrapped.f$ becomes a private function and cannot be called directly on the blockchain. The only way to call it is through the \texttt{requestCall} and simulation procedure above. This concludes our protocol.

\paragraph{Example} Figure~\ref{fig:ctow} shows the changes applied to each function of the contract $\contract$ of Figure~\ref{fig:excontract}, when it appears in $\wrapped.$  Note that all of these functions become private and can only be called by $\wrapped.\texttt{simulate}.$ Moreover, they all get an index parameter $k$ which is the index in the ledger at which this function call was requested.

\begin{figure}
	\begin{lstlisting}[language=Solidity,mathescape=true]
function L_start(uint k, bytes32 a) private {
	b[$\contract$] += get_msg_value(k);
	b[get_msg_sender(k)] -= get_msg_value(k);
	require(get_msg_sender(k)==owner);
	desiredResult=a;
}			

function L_getReward(uint k, uint y) private {
	uint x = get_block_number(k);
	address payable recipient = payable(get_msg_sender(k));
	if(keccak256(abi.encodePacked(x+y))==desiredResult) {
		b[recipient] += b[$\contract$];
		b[$\contract$] -= b[$\contract$];
	}
}	
		
function L_cancel(int k) private {
	require(get_msg_sender(k)==owner);
	b[get_msg_sender(k)] += b[$\contract$];
	b[$\contract$] -= b[$\contract$];
}
	\end{lstlisting}
	\caption{Changes to the Functions of the Original Contract $\contract$ of Figure~\ref{fig:excontract} in the Lazy Contract $\wrapped$.}
	\label{fig:ctow}
\end{figure}
\chapter{Experimental Results} \label{sec:exper}

\section*{Implementation}
We implemented our approach, i.e.~an automated tool to obtain the lazy contract $\wrapped$ from any given contract $\contract$ in Python 3. We used Slither~\cite{DBLP:conf/icse/FeistGG19}, Web3Py~\cite{web3py} and Hardhat~\cite{hardhat} for parsing smart contracts written in Solidity and simulating the gas usage on our machine. The implementation is released as open-source software.

\section*{Benchmarks and Experimental Setting} 
As benchmarks, we took verified real-world smart contracts available in the Etherscan database~\cite{etherscan} that were deployed between \todo{January 1, 2022, 00:00:00 UTC and June 30, 2022, 23:59:59 UTC}. Note that this is a small fraction of the contracts deployed on the Ethereum network, but the Solidity source of other contracts are not available to us to experiment with. We then limited our analysis to \todo{\textbf{160,735}} contracts. These were all the available contracts to which our parser and simulator were applicable. Note that some contracts are written in other languages or in older versions of Solidity and require a different parser, and were hence excluded. Thus, we report results over a total of \todo{\textbf{160,735}} real-world contracts.

For each contract $\contract$ in this list, we downloaded the function calls to $\contract$ that were registered on the Ethereum blockchain from the time it was deployed to \todo{31 January 2023, 23:59:59 UTC}. We thus gathered a total of \todo{\textbf{9,055,492}} transactions. We then applied our protocol and performed the exact same function call requests in the exact same order, environment, and gas price for all of our benchmarks. Finally, we compared the total gas usage incurred by our method with the actual total that was paid on the Ethereum blockchain by real-world users.  

\subsection*{Machine and Runtimes} All computations were performed on an Intel Xeon Gold 5115 CPU (2.40GHz, 8 cores) running Ubuntu 20.04 and 16 GB of RAM. Our average runtime for processing and wrapping a smart contract $\contract$ was \todo{3.47s}. This shows our approach is scalable and easily applicable to any real-world contract.

\section*{Overall Savings in Gas Usage} The total gas usage was reduced from \todo{1,459,716,608,439 gas units $\approx$ 75,368.13 ETH $\approx$ {199,755,699 USD} to 651,027,233,021 gas units $\approx$ 34,092.55 ETH $\approx$ {89,843,820} USD}\footnote{We do not use a fixed exchange rate for ETH/USD. The conversions are done at the time of the actual transactions according to data provided by Etherscan~\cite{etherscan}. Similarly, the conversion between gas units and ETH also has a variable rate based on the gas price used in the original transaction.}. So, our approach provided a gas usage reduction of \todo{\textbf{55.4 percent}}, corresponding to an astonishing \todo{\textbf{109,911,879 USD}}. Each contract's gas usage was, on average, reduced by \todo{\textbf{56.01 percent}}. This shows the significant real-world utility of our method, as well as the fact that current Ethereum smart contracts are quite wasteful in terms of gas usage. 

\subsection*{Detailed Results} 
\begin{sidewaysfigure}
	\includegraphics[keepaspectratio,width=\linewidth]{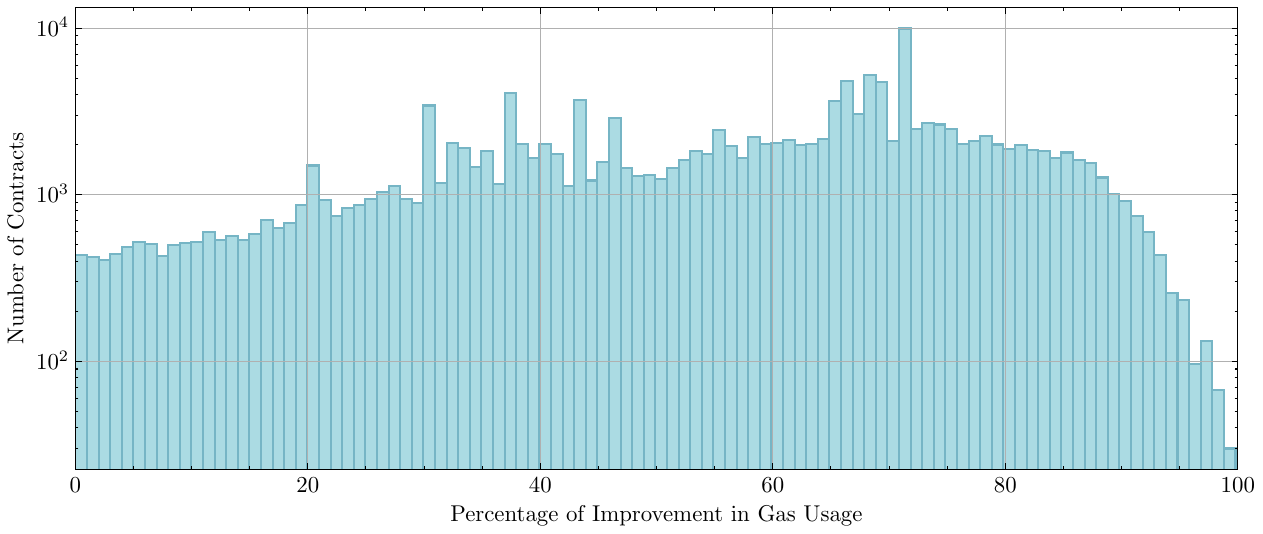}
	\caption{Percentage of Gas Usage Improvements for Benchmark Contracts (Based on Gas Units).}
	\label{fig:gas_saved_percent_per_contract_hist}
\end{sidewaysfigure}

\begin{sidewaysfigure}
	\includegraphics[keepaspectratio,width=\linewidth]{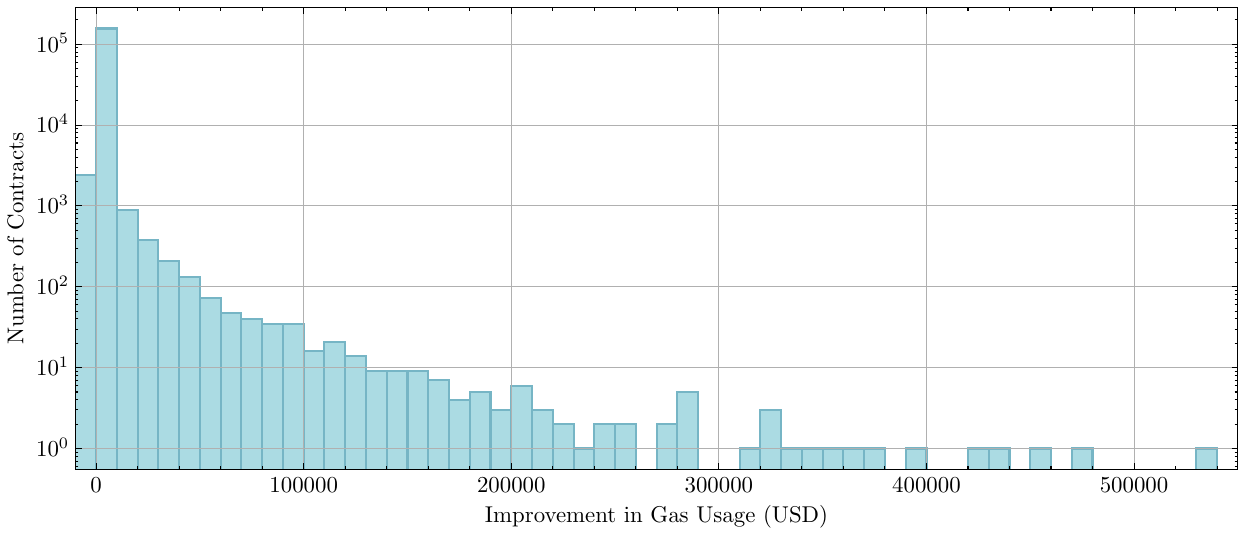}
	\caption{Cost Improvements for Benchmark Contracts (in USD).}
	\label{fig:USD_saved_per_vontracts_hist}
\end{sidewaysfigure}

Figures~\ref{fig:gas_saved_percent_per_contract_hist} and \ref{fig:USD_saved_per_vontracts_hist} provide more detailed results.
Figure~\ref{fig:gas_saved_percent_per_contract_hist} shows the percentage of improvement in gas usage per contract with a histogram bin size of $1\%$, whereas
Figure~\ref{fig:USD_saved_per_vontracts_hist} provides the USD value of the improvements per contract. In Figure~\ref{fig:USD_saved_per_vontracts_hist}, the histogram bin size is 10,000 USD. The x-axis is cut at 550,000, which excludes 7 data points (contracts). On these contracts, our approach saved a total of 3,596,553 USD. Note that the $y$-axis is in logarithmic scale. Due to the variable exchange rate, our approach loses USD on some contracts despite saving gas units.

\subsection*{Discussion} As shown by our experiments, our approach reduces the gas usage of current real-world Ethereum smart contracts by a significant margin of \todo{55.4 percent}. Still, these figures fail to capture the true potential of our method. Many real-world smart contracts are currently simple and have small gas usage precisely due to the limitation of high gas costs. Our approach lifts this restriction and allows smart contract programmers to confidently develop more complicated contracts that require more gas, since the actual incurred gas in our approach is going to remain $O(1)$ for every function call, no matter how gas-consuming the original call. On Ethereum alone, our method saves hundreds of millions of dollars of gas usage per year. These savings will only become more pronounced as time goes by and both the ether's value and the adoption of smart contracts increase. Moreover, our approach is applicable to any other programmable blockchain, too.

\chapter{Extensions, Limitations and Threats to Validity}
\label{sec:lim}

We now discuss the assumptions and limitations behind our approach, as well as possible extensions and enhancements. Note that the limitations mentioned below are also present in other gas-saving measures and related works. See Chapter~\ref{ch:related}.

\section*{Multi-contract Wrappers} 
While we explained our process for wrapping only a single contract $\contract,$ this choice was for ease of presentation. In practice, a single wrapper lazy contract $\wrapped$ can cover many original contracts $\contract_1, \ldots, \contract_n$ and can also handle inter-contract function calls between them. This is also supported by our implementation in Chapter~\ref{sec:exper}. However, the contracts wrapped in $\wrapped$ are essentially disconnected from the rest of the blockchain and our approach cannot support inter-contract function calls between the $\contract_i$'s and other external contracts. This limitation is inherent and cannot be lifted given that our protocol performs lazy function calls, which is not supported in the rest of the blockchain.

\section*{Active Participation} 
Our protocol depends on the active participation of at least one of the parties in $\parties$ in off-chain execution. If $\alice$ tries to commit fraud and withdraw more than her $\wrapped$-balance, the only way to stop her is if another party $\bob$ issues a challenge on time. Therefore, at least one other party must monitor the blockchain and create challenges when necessary. However, this limitation is not a significant issue as most real-world scenarios involve all parties monitoring the blockchain. Also, to further incentivize the issuance of challenges, we can edit $\wrapped$ to penalize a dishonest $\alice$ by more than the actual on-chain gas cost and pay the remainder to the first party $\bob$ who challenges the fraudulent withdrawal. This ensures that even non-parties have an incentive to join and report fraudulent behavior. The same idea can be applied to further incentivize the initiators and pay more fees to $\ingrid.$

\section*{Delay-of-Withdrawal Attack} 
In our protocol, a malicious party $\bob$ can attempt to stop a valid withdrawal by $\alice$ by challenging it. However, this challenge can only delay the withdrawal as it will cause an on-chain execution which uncovers $\bob$'s dishonesty. Since all the gas cost is paid by $\bob,$ there is no financial downside for $\alice$ in this case. We can also add a further penalty for a dishonest $\bob$ and use it to compensate $\alice$ for the delay she faces. However, in practice, having to pay for the entire contract's gas usage is usually a large enough disincentive to dissuade such behavior.

\section*{Gas Usage} 
Assuming all parties are rational and therefore honest, our protocol will never cause an on-chain simulation. Hence, its gas usage is mainly due to the records added to the $\wrapped$-ledger and each function call request incurs at most a constant $O(1)$ amount of gas. For most real-world contracts, this means $\wrapped$ uses significantly less gas than $\contract.$ However, if $\contract$ is itself very simple, e.g.~an ERC20 token or a simple NFT with no extra functionality, in which every function call already takes $O(1)$ gas, then applying our protocol would have no benefit and might even slightly increase the gas usage.

\subsection*{Stored Global Variables} We mentioned that for every function call request, we add the values of all global variables to our ledger. In practice, our implementation is optimized and performs a syntactic analysis that ensures only those global variables whose values are needed are added to the ledger. This helps reduce the amount of storage used for the $\wrapped$-ledger.

\section*{Deposits} 
Finally, the main drawback of our approach is that every participant has to pay a deposit $d$ when joining the contract and this deposit should be large enough to cover the potential gas costs in any case. While we have no way of avoiding this deposit, since on-chain executions are unpredictable in advance and have to be billed to the dishonest party, note that the deposit is reimbursed when the party leaves the contract. Thus, we argue that having to put down a deposit that is reimbursed to every honest participant is a much more attractive proposition than having to pay higher gas fees that are never paid back.

\section*{Checkpointing} If a contract is expected to handle many function calls, then the deposit amount $d$ can grow large since it should be enough to cover an on-chain execution. To avoid such a scenario, our approach can be easily extended to incorporate checkpointing. More specifically, suppose our goal is to ensure that the on-chain execution, if it is ever triggered, will not have to simulate more than $k$ function calls. We can require the parties to add a checkpoint before every $k$-th function call\footnote{More specifically, a party will not be able to add the $k$-th function call request unless she adds a checkpoint first.}. A checkpoint is simply a record which contains the hash of the current state of the contract's storage and the balances in the offline eagerly-executed version. Thus, it is a very short digest and adding it takes $O(1)$ gas. Other parties can see the checkpoint and, if it does not match their current off-chain state, they can challenge it in the exact same manner as a withdrawal is challenged. If a checkpoint is not challenged within $t$ blocks, it stands. Otherwise, an on-chain execution will be triggered. The upside of this approach is that the on-chain execution should only simulate the function calls since the last checkpoint. In other words, $\ingrid$ first provides the state of the balances and $\contract$'s storage at the time of the last checkpoint. The hash of this state is checked by $\wrapped$ to ensure $\ingrid$ is being honest. Then, only the function calls since the last checkpoint will be simulated on-chain in the same manner as before.

\chapter{Related Works}\label{ch:related}

Prior works on reducing the gas usage of smart contracts can be divided into three general categories: (i)~layer-one improvements, (ii)~approaches based on layer-two protocols, and (iii)~static analysis. Below, we provide an overview of each category.

\section*{Layer-1 Solutions}
In both academia and industry, there has been a variety of suggestions as to how blockchains can be scaled without sacrificing their decentralization or security characteristics. Most such solutions are generally referred to as Layer-1 approaches, as they make changes to how the underlying blockchain works and require a hard fork. These solutions can mainly be categorized into three types~\cite{gangwal2023survey}:

\begin{itemize}
    \item \emph{Alternative consensus mechanisms:} Bitcoin uses a permissionless consensus protocol that involves cryptographic block-discovery races that exhaustively probe cryptographic hash functions to generate partial preimages. This is called Proof of Work (PoW)~\cite{nakamoto}. Practical PoW-based blockchain networks, such as Bitcoin, have witnessed a rapid increase in their total hash rate which negatively impacts their transaction throughput and transaction fees~\cite{wang2019survey}. Therefore, in order to improve the scalability of PoW, new consensus algorithms have been proposed such as Proof of Stake (PoS)~\cite{DBLP:conf/crypto/KiayiasRDO17}, Hybrid Mining~\cite{DBLP:conf/sac/ChatterjeeGP19}, and Proof of Space~\cite{DBLP:conf/fc/ParkKFGAP18}.
    \item \emph{Sharding:} Parallel processing of transactions, as suggested in sharding-based blockchain protocols, can increase transaction processing power as more users join the network. As a result, the network is capable of handling high transaction loads and scaling effectively~\cite{yu2020survey}. There have been many works on sharding-based protocols in recent years~\cite{kokoris2018omniledger, gencer2016service}. For example, the Elastico protocol separates the mining network into smaller committees, each of which processes its own transaction set. This results in practically linear scaling of transaction rates with available mining power~\cite{luu2016secure} and RapidChain uses an efficient cross-shard transaction verification technique to eliminate the need for gossiping transactions across the entire network~\cite{zamani2018rapidchain}.
    \item \emph{Modifying block data:} This research direction focuses on optimizing blocks to achieve a general bandwidth reduction in the network when blocks or transactions are shared~\cite{antwi2022survey}. An example of this paradigm is LightBlock, which replaces the transactions in the original block with the transaction hash, resulting in a much smaller bandwidth resource usage than the propagation of the original block~\cite{LightBlock}.
\end{itemize}

Since Layer-1 solutions alter the fundamental structure of blockchains, such as the consensus mechanism and network communication, they lack backward-compatibility and require hard forks. Therefore, such upgrades are not practical and are rarely applied in practice. Ethereum's roadmap suggests that sharding is no longer required for scaling since layer-2 rollups are a successful alternative~\cite{ethereum_roadmap}. The advantages of our lazy contracts over layer-1 solutions are that (i)~they execute exactly as any other smart contract, and (ii)~they do not modify the fundamental properties of the blockchain, or require a hard fork, making them more practical.

\section*{Layer-2 Solutions}
Security and decentralization, which are fundamental properties of every blockchain system, should not be adversely affected by scaling improvements. Researchers have been studying ways to create a scalable blockchain by building what is known as a second layer. The goal is to have another blockchain (layer-2) on the top of the original blockchain, which is referred to as main-chain or layer-~\cite{sguanci2021layer}. The design of layer-2 protocols can vary, but their primary function is to handle transactions off-chain (off the layer-1 chain) and to only use the main chain when resolving disputes or reporting summaries. This will in turn reduce the load on the layer-1 blockchain itself, resulting in a lower transaction fee through lower gas usage. We provide a summary of prominent layer-2 solutions following~\cite{gangwal2023survey}.

\subsection*{Channels}
Channels are one of the most commonly-used proposals for addressing the scalability challenges in cryptocurrencies. Initially, channels focused on facilitating secure off-chain payments between two parties, and were referred to as payment channels. A payment channel consists of two parties agreeing to an initial state on-chain, and once the transaction has been mined, the parties will be able to pay each other securely by updating the channel balance. As with any layer-2 protocol, on-chain communication will only be employed when one of the parties attempts to close the channel or punish the other for dishonest behavior~\cite{DBLP:conf/ccs/DziembowskiFH18}. Payment networks, which allow users to route transactions through intermediaries, are an important extension to payment channels. Moreover, payment channels have also been generalized so they can execute more complex smart contracts. These generalizations are known as state channels. Examples of these approaches are~\cite{poon2016bitcoin,raiden,DBLP:conf/ccs/DziembowskiFH18,DBLP:conf/eurocrypt/DziembowskiEFHH19}. In contrast to our approach, state channels are limited in the types of contracts they can handle and are not applicable to arbitrary smart contracts written in a Turing-complete language such as Solidity.

\subsection*{Cross Chains}
Cross-blockchain communication involves two blockchains: a source blockchain and a target blockchain. Source blockchains are the blockchains from which transactions are initiated and executed on target blockchains. A cross-chain monetary exchange is a special case of cross-chain communication that has been researched extensively and also implemented for real-world usage. In general, cross-chain transactions consist of three steps: (i) Committing to assets on the source blockchain; (ii) transferring commitments; and (iii) representing equal assets on the target blockchain~\cite{belchior2021survey}. Examples of cross-chain communications are~\cite{tian2021enabling, zamyatin2019xclaim}. Cross-chain protocols can be used to transfer the execution of the main contract to a cheaper blockchain, while taking deposits in a more commonly-used blockchain, such as Ethereum. The downsides are that the programmer has to write contracts for both blockchains and the contract-executing blockchain might not be as secure as the original one.

\subsection*{Side-chains and Rollups}
A side-chain is an independent distributed ledger that runs in parallel to the main chain and serves to reduce the load on the main chain by transferring computationally heavy processes off-chain. Side-chains may have different communication protocols or consensus mechanisms than their main-chain. There are two prominent types of side-chains: custodial and non-custodial. Custodial side-chains rely on their own consensus mechanisms, while non-custodial side-chains secure their state on the main chain~\cite{gangwal2023survey}. Rollups are currently the most practical and promising side-chain design approach, and will be discussed in greater detail below. They are also the closest approach to lazy contracts in terms of their design.

Ethereum proposed rollups in 2018. Since then, they have been characterized as a hybrid solution between layer-1 and layer-2 solutions. This is because, while layer-2 solutions, such as channels, update the on-chain state with the final result, rollups maintain a record of every single transaction that took place off-chain within the on-chain contract. Specifically, in rollups, a smart contract deployed on-chain knows the latest (current) Merkle root of the rollup's state, thus eliminating data withholding issues, because all the information is always retrievable on-chain~\cite{sguanci2021layer}. Various actors in rollup systems interact with this smart contract, including sequencers, which after executing transactions publish the compressed data (i.e. rollup data) on-chain, as well as verifiers, who dispute transactions when necessary. There are two kinds of rollups: optimistic rollups and zero-knowledge (ZK) rollups.

\subsubsection{Optimistic Rollups} Optimistic rollups involve an aggregator gathering data off-chain, executing transactions off-chain, and updating the new state. Verifiers continuously monitor published new states, and if a violation is detected, they initiate a dispute phase. The verifier awards a portion of an aggregator's deposit if the dispute phase shows malicious behavior. Aggregators are therefore incentivized to act honestly to earn block rewards and not lose their deposits, while verifiers need to process new states to earn a reward if they find malicious behavior.
\begin{itemize}
    \item \textbf{Optimism:} Optimism~\cite{optimismworks} is an optimistic rollup on the top of Ethereum that uses a virtual machine called Optimism Virtual Machine (OVM) which is compatible with the Ethereum Virtual Machine (EVM). By leveraging an on-chain smart contract $c_o$, optimism allows its sequencers to publish new rollups on-chain and verifiers to challenge its transactions if they detect fraudulent ones. At the time of writing, the Optimism Foundation runs the only sequencer. Thus, the system is highly centralized. Suppose that a smart contract $c$ is deployed off-chain and recent rollups contain a transaction $t$ that is called a function of $c$. Given that the transaction data in $c_o$ is compressed, the verifier needs to provide sufficient data in $c_o$ to specify that $t$ needs to be executed. Then, the verifier deploys $c$ on-chain and asks $c_o$ to verify it. Next, the verifier sets the recent state of $c$ on-chain and asks $c_o$ to execute it. Then, $c_o$ checks if the new state of $c$ matches the one that the sequencers submitted, and punishes dishonest parties accordingly~\cite{thibault2022blockchain}. This is similar to our challenge phase, except that in our system, anyone can take the roles of $\bob$ and $\ingrid$ and there is no centralization.

    \item \textbf{Arbitrum:} Offchain Labs has been developing Arbitrum~\cite{arbitrumworks} since 2018 on top of Ethereum. This is a layer-2 network based on optimistic rollups that aims to minimize computation on the Ethereum network during dispute resolutions. Arbitrum uses its own Arbitrum Virtual Machine (AVM) to execute transactions and resolve disputes using a one-step proof. In addition, Arbitrum incorporates an on-chain smart contract $c_a$, which allows the aggregator to publish new rollups, and the verifier to check whether they are fraudulent. As in the previous case, at the time of writing, Offchain Labs is the only sequencer operating on Arbitrum. For a better understanding of the dispute phase, suppose that the smart contract $c$ is deployed off-chain and the most recent rollup contains a transaction $t$ that invoked a function of $c$. The Arbitrum verifier must not only report the invalid transaction to $c_a$ but also identify which opcode $op$ of the transaction was executed incorrectly. Then, since AVM is a stack-based virtual machine, the verifier only has to set the stack state before $op$ so that $c_a$ can execute AVM for $op$. Lastly, $c_a$ compares the new AVM state (as opposed to optimism, which compares the contract state) to what the aggregator submitted and punishes dishonest behavior accordingly~\cite{thibault2022blockchain}.
\end{itemize}

\subsubsection{ZK Rollups}
Much like optimistic rollups, ZK rollups allow parties to process transactions off-chain and publish the new state on-chain in batches. Contrary to optimistic rollups, which assume that transactions are valid until objected to by a verifier, in ZK rollups, the sequencers must provide a cryptographic zero-knowledge proof that transactions are correct. As on-chain computations are expensive, the proof size and verification time should be small and grow slowly as the number of off-chain transactions increases. Currently, these protocols use popular non-interactive proofs such as SNARKs, STARKs and Bulletproof~\cite{zksurvey}. ZKSync~\cite{zkSync} and STARKEx~\cite{starkex} are examples of such approaches.

\subsubsection{Comparison with Our Approach}

Since our protocol shares several design patterns with optimistic rollups, we now further compare lazy contracts with these approaches. Unlike our approach, rollups have their own mining protocol, consensus mechanism, and implementation separate from the main blockchain. They have few active miners and hence the side-chain is highly centralized. Moreover, they often have central authority figures and are sometimes not even trustless. For instance, the sequencer is currently a single centralized entity in both Optimism~\cite{optimismworks} and Arbitrum~\cite{arbitrumworks}. Our approach does not require a separate blockchain and is as decentralized as the layer-one blockchain. It is also completely trustless. Additionally, a developer who uses a sidechain has to put in extra effort and consider both layers, which usually have different environments and programming languages or bytecode formats.

In addition to the points and differences mentioned above, note that layer-two solutions and rollups similarly share all the limitations of our approach which were discussed in Chapter~\ref{sec:lim}. They can handle multiple contracts at once, but their contracts form a bubble and are unable to communicate with the rest of the blockchain. Participants must also actively challenge dishonest behavior by other users and lock a deposit. Finally, they have time-consuming dispute resolution mechanisms, which opens the possibility of delay-of-withdrawal attacks. Therefore, the popularity of channels and optimistic rollups leads us to believe that such limitations will be tolerated by developers in exchange for the significant reduction in gas costs. Indeed, switching to our approach is simpler than using rollups and does not require any additional development effort since the wrapping procedure is entirely automated and push-button.

\section*{Static Analysis} 
Static analysis of smart contracts is another well-studied topic that is not only focused on gas optimization but also considers vulnerability detection~\cite{DBLP:conf/kbse/0001LC18,DBLP:journals/pacmpl/GrechKJBSS18,DBLP:journals/fgcs/LiJCLW20}. Many previous works rely on static analysis to optimize the source code or bytecode of a contract before its deployment and hence reduce its gas usage. The main idea is to keep the contract's functionality unaffected, i.e.~preserve its semantics, while avoiding costly gas patterns. Some works in this direction are as follows:
\begin{itemize}
	\item The work \cite{MarchesiDesign} provides 24 design patterns to reduce gas usage in developing Ethereum smart contracts. 
	
	\item GasMet~\cite{DBLP:journals/jss/SorboLVVC22} identifies 19 costly code smells, i.e.~patterns that lead to higher gas usage. It does not automatically optimize the code but instead issues warnings to the programmers and points them to the occurrence of the pattern.
	
	\item GasReducer~\cite{DBLP:conf/icse/ChenLZCLLZ18} automatically detects and replaces anti-patterns and thus performs bytecode optimization. Similarly, GasSaver~\cite{DBLP:conf/bsci/NguyenDND22} optimizate the source code.
	
	\item  GASPER~\cite{chen2017under} is a tool for detecting gas-based vulnerabilities in Ethereum smart contracts using symbolic execution. It is extended by GasChecker~\cite{DBLP:journals/tetc/ChenFLZLLXCZ21} to detect ten costly patterns in four categories: (i)~useless code, (ii)~loops, (iii)~wasted storage, and (iv)~gas-inefficient operation sequences. 
	\item The work~\cite{DBLP:conf/cav/AlbertGRS20} uses SMT-based super-optimization and attempts to find the best equivalent block of code for each basic block of the original smart contract.
	\item The work \cite{DBLP:journals/jcst/KongWHCZZH22} focuses on static optimization of gas costs at source-code level instead of bytecode. This is achieved by a survey of online forum discussions that manually identified common causes of gas inefficiency in Solidity.
	\item In~\cite{DBLP:conf/brains/NelaturuBLV21}, a synthesis-based approach is used to optimize and summarize loops in smart contracts.
	\item GASOL~\cite{DBLP:conf/tacas/AlbertCGRR20} also uses static analysis. In contrast to other methods, it also has a focus on storage optimization.
 \end{itemize}

\subsection*{Comparison with Our Approach}

 All static analysis approaches above keep the core functionality of the contract intact and keep it entirely on-chain. So, they can only obtain small reductions in overall gas usage. In contrast, we move all of the execution off-chain and sidestep the problem of costly on-chain computations entirely. This leads to much larger improvements and only a small constant amount of gas usage per function call. A direct empirical comparison is impossible since most of the tools above are not publicly available and the rest are designed for older versions of Solidity. However, the improvements obtained by our approach are two orders of magnitude larger. For example, according to their own experiments, GasReducer saves 1,525 units of gas per transaction, GasChecker 22 units, and~\cite{DBLP:journals/jcst/KongWHCZZH22} saves 245 units. In contrast, we save \todo{$\sim$89,300 units} per transaction on average and our median saving per transaction is 29,390 gas units.

\chapter{Conclusion}

In this thesis, we provided a protocol based on lazy on-chain execution of function calls to reduce the gas usage of smart contracts. The central idea in our method is to be lazy when it comes to on-chain computations but eager with off-chain execution of the contract. We implemented our approach and showed that, on the Ethereum blockchain alone, it saves hundreds of millions of dollars per year in gas costs. Additionally, our approach is trustless and decentralized and thus preferable to current state-of-the-art layer-two solutions. It is truly remarkable that a well-known programming languages concept, i.e.~lazy evaluation/execution, can be integrated in the context of blockchain and enable such huge real-world savings.

\nocite{farokhnia2023alleviating,farokhnia2023reducing,cai2023game,cai2023trustless}

\newpage
\addcontentsline{toc}{chapter}{References}
\bibliographystyle{IEEEtranN}
\bibliography{reference} 

\newpage
\addcontentsline{toc}{chapter}{Publications}
\null\skip0.2in
\begin{center}
{\bf \Large \underline{List of Publications}}
\end{center}
\vspace{12mm}

\begin{itemize}
    \item S. Farokhnia and A.K. Goharshady, ``Alleviating High Gas Costs by Secure and Trustless Off-chain Execution of Smart Contracts,'' in Proceedings of the 38th ACM/SIGAPP Symposium on Applied Computing (SAC), 2023
    \item S. Farokhnia and A.K. Goharshady, ``Reducing the Gas Usage of Ethereum Smart Contracts without a Sidechain,'' in Proceedings of the 5th IEEE International Conference on Blockchain and Cryptocurrency (ICBC), 2023
    \item Z. Cai, S. Farokhnia, A. Goharshady, S. Hitarth, ``Asparagus: Automated Synthesis of Parametric Gas Upper-bounds for Smart Contracts,'' in Proceedings of the ACM SIGPLAN International Conference on Object-Oriented Programming Systems, Languages, and Applications (OOPSLA), 2023
\end{itemize}


\end{document}